\documentclass[journal, twocolumn, 10pt]{IEEEtran}

\usepackage[pdftex]{graphicx}
\usepackage[cmex10]{amsmath,empheq}
\usepackage{cite}
\usepackage{url}
\usepackage{color}
\usepackage{microtype}
\usepackage{amssymb}
\usepackage{relsize}
\usepackage{lipsum}
\usepackage{epstopdf}
\usepackage[font = footnotesize]{caption}
\usepackage{enumitem}
\usepackage{bm} % (Italic) Bold math
\usepackage{multirow}
\usepackage{array}
\usepackage{ifthen}
\usepackage{float}
\usepackage{subfig}
\usepackage{xcolor}% http://ctan.org/pkg/xcolor
\usepackage{algpseudocode}%http://ctan.org/pkg/algorithmicx
\usepackage{algorithm}% http://ctan.org/pkg/algorithm
\usepackage{soul}

\definecolor{pblue}{HTML}{0072BD}

\DeclareMathOperator{\VEC}{vec}
\DeclareMathOperator{\diag}{diag}
\usepackage{mathptmx} % IEEE prefers to use times also for math
\DeclareMathAlphabet{\mathcal}{OMS}{cmsy}{m}{n}

\usepackage{siunitx} % For typesetting values with units
\algnewcommand{\algorithmicand}{\textbf{ and }}
\algnewcommand{\algorithmicor}{\textbf{ or }}
\algnewcommand{\OR}{\algorithmicor}
\algnewcommand{\AND}{\algorithmicand}
\algnewcommand{\var}{\texttt}

\begin{document}

\title{PAPR Reduction with Mixed-Numerology OFDM}

\author{Selahattin G\"{o}kceli,
		Toni Levanen,
		Juha Yli-Kaakinen,
		Taneli Riihonen,
		Markku Renfors,
        and~Mikko Valkama 

\thanks{S. G\"{o}kceli, T. Levanen, J. Yli-Kaakinen, T. Riihonen, M. Renfors, and M. Valkama are with 
Tampere University, Tampere, Finland (\emph{Corresponding Author:} Mikko Valkama, e-mail: mikko.valkama@tuni.fi).

© 2019 IEEE.  Personal use of this material is permitted.  Permission from IEEE must be obtained for all other uses, in any current or future media, including reprinting/republishing this material for advertising or promotional purposes, creating new collective works, for resale or redistribution to servers or lists, or reuse of any copyrighted component of this work in other works.

This work has been accepted for publication as a Letter in the IEEE Wireless Communications Letters. Copyright may be transferred without notice, after which this version may no longer be accessible.}}

\maketitle

\begin{abstract}
High peak-to-average power ratio (PAPR) is a critical problem in orthogonal frequency-division multiplexing (OFDM). The fifth-generation New Radio (5G NR) facilitates the utilization of multiple heterogeneous bandwidth parts (BWPs), which complicates the PAPR problem even further and introduces inter-numerology interference (INI) between the BWPs. This paper proposes two novel schemes to reduce the PAPR of mixed-numerology OFDM signals. The first scheme is an original enhanced iterative clipping-and-error-filtering (ICEF) approach that cancels efficiently the INI along with PAPR reduction. This allows to achieve efficient PAPR reduction while being compatible with well-known windowed overlap-and-add (WOLA) processing. The second scheme is based on fast-convolution (FC) processing, where PAPR reduction is embedded in the FC filtering carried out using overlapping processing blocks. This allows one to use any existing block-wise PAPR reduction method to reduce the composite signals' PAPR with arbitrary BWP waveforms, and it is thus especially well suited for processing mixed-numerology composite waveforms carrying multiple BWPs with different OFDM numerologies. The performance of the proposed algorithms is evaluated in the 5G NR context, and the essential performance advantages are demonstrated and quantified.  
\end{abstract}
\vspace{-1mm}

\section{Introduction}
\label{sec:intro}

\IEEEPARstart{F}{ifth-generation} New Radio (5G NR) promises dramatic improvements over existing systems in terms of data rates, reliability, latency, and energy consumption \cite{DAHLMAN201857,NR.300}. The physical-layer radio access of 5G NR is based on cyclic-prefix (CP) orthogonal frequency-division multiplexing (OFDM), while it also facilitates simultaneous utilization of non-orthogonal subbands---or bandwidth parts (BWPs) in 3GPP terminology---with different numerologies and quality-of-service (QoS) expectations \cite{DAHLMAN201857,NR.300}. To reduce the peak-to-average power ratio (PAPR), the clipping of different BWPs could be done separately. However, as will be shown in this paper, such approach does not facilitate efficient PAPR reduction of the aggregated signal combining multiple BWPs. Additionally, in CP-OFDM systems supporting different numerologies, inter-numerology interference (INI) is introduced \cite{DAHLMAN201857,fc2}. Therefore, the mixed-numerology signal structure and possible increase of INI should be considered when designing and developing the PAPR reduction algorithms. 

The well-known PAPR reduction methods for single-numerology OFDM, such as iterative clipping and filtering (ICF) \cite{clipping}, partial transmit sequence \cite{Jawhar}, selected mapping \cite{2010LiSLM}, and tone reservation \cite{7024183} do not consider INI and are thus limited in performance. Different windowing- or filtering-based spectrum enhancement techniques, in turn, are widely considered for suppressing INI, but do not consider the PAPR problem. Thus, the spectrum enhancement and PAPR reduction processes should be jointly considered, something that is addressed in this paper. Overall, mixed-numerology OFDM signal PAPR reduction has received surprisingly small attention in the related literature, increasing the importance of this work. 

The windowed overlap-and-add (WOLA) processing \cite{2013:J:Bala:WOLARef} is, in general, a well-known solution to reduce INI. In this paper, we present a novel iterative-clipping-and-error-filtering (ICEF)-type algorithm applied on OFDM symbol basis for mixed-numerology waveforms. Traditional ICF \cite{clipping} and ICEF \cite{icef} solutions suffer from the INI, leading to very poor PAPR reduction performance. The proposed enhanced ICEF (E-ICEF) relies on cancelling the INI in each BWP during the iterative PAPR reduction process. This ensures significantly improved PAPR reduction performance and convergence of the PAPR reduction algorithm. After the E-ICEF processing of the mixed-numerology signal, WOLA is applied to each BWP component before aggregating them into the final transmitted signal.

Filtered OFDM is another effective approach for INI mitigation, and fast-convolution filtered OFDM (FC-F-OFDM) \cite{fc2} is a computationally efficient and very flexible frequency-domain method for mixed-numerology systems, like 5G NR. In this paper, an effective PAPR reduction scheme is proposed by embedding properly tailored ICEF-like processing into the FC filtering. The novelty of the proposed scheme is the ability to jointly mitigate INI, reduce PAPR efficiently, and provide good spectral localization. Since FC-based filtering inherently processes the incoming BWP signals in overlapping blocks, it allows to use any block-based PAPR reduction algorithm to any input signal, irrespective of its more specific nature. The proposed scheme is especially suitable for mixed-numerology scenarios, as it provides computationally efficient PAPR reduction combined with good INI suppression.

\section{System Model and Proposed Solutions}

\subsection{Reference Subband-based Independent ICEF (I-ICEF)}

In the simplest form, PAPR reduction can be applied separately on each subband's CP-OFDM signal, after which these signals are added together to obtain the aggregated mixed-numerology waveform. This scheme is used as a reference, and the baseline ICEF algorithm \cite{icef} is applied as the PAPR reduction method in each subband, such that the clipping noise is allowed only within the active subcarriers of each subband. This prevents the mean squared error (MSE) degradation on other subbands due to the iterative PAPR reduction. To this end, with $M$ subbands, the processed signal can be defined as
\begin{equation}
\mathbf{y}_{\textrm{t}}^\textrm{I-ICEF} =  \sum_{m=0}^{M-1}  \mathbf{K}_m \VEC \left( \textbf{W}^{-1}_{\text{OFDM},m} f_{\textrm{ICEF}}(\textbf{X}_{\textrm{f},m}) \right),
  \label{eq:referenceMethod}
\end{equation}
where $m$ denotes the subband index, $\textbf{W}^{-1}_{\textrm{OFDM},m}$ and $\textbf{X}_{\textrm{f},m}$ represent the $L_{\text{OFDM},m}\times L_{\text{OFDM},m}$ inverse discrete Fourier transform (IDFT) matrix and $L_{\text{OFDM},m} \times S_m$ frequency-domain data matrix (including zero padding) for $S_m$ OFDM symbols, respectively. Here, $L_{\text{OFDM},m}$ represents the nominal transform size of the OFDM processing that is applied for the $m\textrm{th}$ subband. We note that for clarity, the subindices $\textrm{t}$ and $\textrm{f}$ are used to differentiate between time- and frequency-domain signals, respectively, throughout this letter. The parallel-to-serial conversion operation is denoted by  $\VEC(\cdot)$, vertically stacking the columns of the input matrix. The ICEF-based PAPR reduction algorithm, defined in \cite{icef}, is denoted as $f_\text{ICEF}(\cdot)$. In the transmitter's WOLA processing, the time-domain OFDM symbols are extended by a CP with $L_{\text{EXT},m}$ samples, multiplied sample-wise by the time-domain window function, and concatenated using the overlap-and-add processing. This is denoted by $\mathbf{K}_m$, which is a block-diagonal matrix $\mathbf{K}_m = \diag\left(\mathbf{O}_{m,0},\mathbf{O}_{m,1},\dots,\mathbf{O}_{m,S_m-1}\right)$ with $\mathbf{O}_{m,r}=\mathbf{D}_{\text{TD},m}\mathbf{T}_{\text{EXT},m}$. Here, $\mathbf{T}_{\text{EXT},m}$ is the $L_{\textrm{WOLA},m}\times L_{\text{OFDM},m}$ time-domain cyclic extension matrix, whereas the time-domain windowing matrix $\mathbf{D}_{\text{TD},m}$ of size $L_{\textrm{WOLA},m} \times L_{\textrm{WOLA},m}$ has the time-domain raised-cosine (RC) window weights \cite{2013:J:Bala:WOLARef} on its main diagonal and $L_{\textrm{WOLA},m}=L_{\textrm{OFDM},m}+L_{\textrm{EXT},m}$.

\subsection{Proposed Enhanced ICEF (E-ICEF)}
\label{subsec:E-ICEF}

\label{sec:system_model}
\begin{figure}[tb]
	\centering
	\includegraphics[width=0.6\columnwidth]{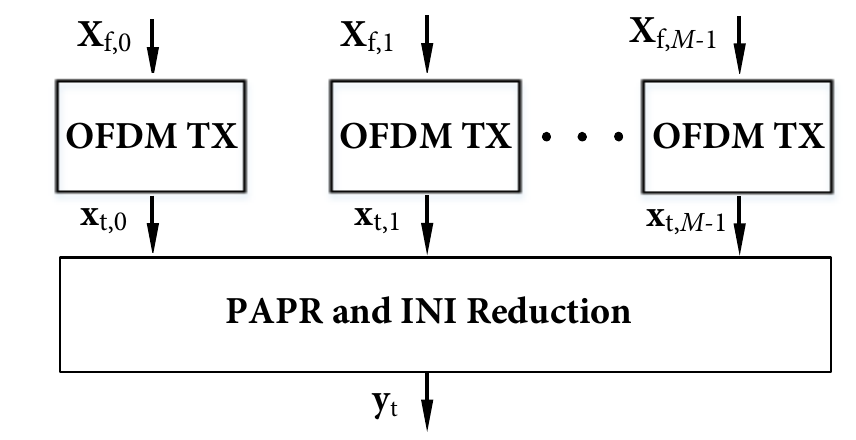}
	\caption{High level block diagram for the proposed mixed-numerology PAPR and INI reduction methods.} 
	\label{fig:PAPRandINI}
\end{figure}

The first proposed method is a novel E-ICEF algorithm, specifically tailored for mixed-numerology operation. In \mbox{E-ICEF}, the PAPR reduction is applied on the aggregated signal, rather than separate input subband signals, as illustrated conceptually in Fig. \ref{fig:PAPRandINI}. This ensures that no new power peaks are generated due to combining of independent signals, and thus clearly improved PAPR performance can be achieved w.r.t.\ \mbox{I-ICEF}. 
However, with OFDM-based mixed-numerology signals, any PAPR reduction algorithm operating in a symbol-wise manner will cause INI, introduced when converting the time-domain clipped OFDM waveform to the frequency domain, as the used subband-specific DFT spreads the interference from the non-orthogonal numerologies to the considered subband. Based on our observations, the level of INI increases in each iteration unless it is controlled during the process. Therefore, to allow the ICEF algorithm to achieve PAPR reduction with reasonable passband MSE, the INI caused by other subbands is properly suppressed in each iteration. 

To this end, assuming $M$ subband signals with subband index $m \in \{0,1,\dots,M-1\}$, the component signals are generated as
\begin{equation}
  \mathbf{x}_{\textrm{t},m} = \VEC \left( \textbf{T}_{\text{CP},m}
  \textbf{W}^{-1}_{\text{OFDM},m} \textbf{X}_{\textrm{f},m} \right),
\end{equation}
where $\textbf{T}_{\textrm{CP},m}$ represents the $(L_{\text{OFDM},m}+L_{\text{CP},m})\times L_{\text{OFDM},m}$ CP insertion matrix. Then, since different subbands are assumed to carry different BWPs using different 5G NR numerologies, the conventional single-transform approach for frequency-domain clipping noise filtering cannot be applied. Instead, $M$ parallel transforms are utilized to correctly filter the clipping noise in each subband. The numerology-specific CPs need to be also removed and regenerated in each iteration, so that the relative timing of different numerology CP-OFDM symbols in different subbands is maintained during the E-ICEF 
process. Accordingly, in the context of iterative PAPR reduction and the $m\textrm{th}$ subband or BWP, clipping noise at $l\textrm{th}$ iteration for OFDM symbol index $s \in \{0, 1, \ldots, S_m-1\}$ can be obtained in frequency domain as
\begin{align}
    \mathbf{c}^{(l)}_{\textrm{f},m,s} = \bar{\mathbf{x}}^{(l)}_{\textrm{f},m,s} - \mathbf{x}_{\textrm{f},m,s} - \mathbf{z}^{(l-1)}_{\textrm{f},m,s},
\end{align} 
where $\bar{\mathbf{x}}^{(l)}_{\textrm{f},m,s}$ is the frequency-domain response of the clipped OFDM symbol at iteration $l \in \{1, 2, \ldots, \mathcal{L}\}$, $\mathbf{x}_{\textrm{f},m,s}$ is the frequency-domain response of the original OFDM symbol, and $\mathbf{z}^{(l-1)}_{\textrm{f},m,s}$ is the INI component at the end of iteration $l-1$ at subband $m$. Importantly, the INI component can be expressed as
\begin{align}
    \mathbf{z}^{(l-1)}_{\textrm{f},m,s} = \textbf{W}_{\textrm{OFDM},m} \sum_{\substack{i=0 \\ i\neq m}}^{M-1} \mathbf{x}^{(l-1)}_{\textrm{t},i,s},
    \label{iniequ}
\end{align}
where $\textbf{W}_{\textrm{OFDM},m}$ is the $L_{\text{OFDM},m}\times L_{\text{OFDM},m}$ discrete Fourier transform (DFT) matrix and $\mathbf{x}^{(l-1)}_{\textrm{t},i,s}$ is an $L_{\textrm{OFDM},m} \times 1$ vector containing time-domain samples with indices $[s(L_{\textrm{OFDM},m}+L_{\textrm{CP},m}), s(L_{\textrm{OFDM},m}+L_{\textrm{CP},m})+1, \ldots, (s+1)(L_{\textrm{OFDM},m}+L_{\textrm{CP},m})-1]$ from the $i{\textrm{th}}$ subband signal after $(l-1){\textrm{th}}$ E-ICEF iteration. 

As shown in (\ref{iniequ}), after the $(l-1)\textrm{th}$ E-ICEF iteration, outputs of all BWPs except the $m\textrm{th}$ one are combined and the aggregated INI on the $m\textrm{th}$ BWP is obtained after taking DFT of size $L_{\textrm{OFDM},m}$. This way, the INI component on the active passband of the evaluated subband is effectively cancelled and the degradation in the MSE and PAPR reduction performance is minimized. After the maximum number of iterations of the E-ICEF algorithm is reached, or the PAPR target is achieved, subband-specific WOLA processing is applied to the PAPR reduced subband signals and they are combined into the mixed-numerology output signal, expressed as
\begin{equation}
  \mathbf{y}_{\textrm{t}}^\textrm{E-ICEF} =  \sum_{m=0}^{M-1}  \mathbf{K}_m \VEC \left( \textbf{W}^{-1}_{\text{OFDM},m} \textbf{X}^{(\mathcal{L})}_{\textrm{f},m} \right),
  \label{eq:E_ICEF_output}
\end{equation}
where $\textbf{X}^{(\mathcal{L})}_{\textrm{f},m} = [\textbf{x}^{(\mathcal{L})}_{\textrm{f},m,0}, \textbf{x}^{(\mathcal{L})}_{\textrm{f},m,1}, \ldots, \textbf{x}^{(\mathcal{L})}_{\textrm{f},m,S_m-1}]$. The vector of subcarrier samples for OFDM symbol $s$ at the processing output is of the form $ \textbf{x}^{(\mathcal{L})}_{\textrm{f},m,s} = \textbf{x}_{\textrm{f},m,s} + \textbf{h}^{\textrm{ICEF}}_{\textrm{f},m} \odot \textbf{c}^{(\mathcal{L})}_{\textrm{f},m,s}$, where $\textbf{h}^{\textrm{ICEF}}_{\textrm{f},m}$ is the clipping error filter response for subband $m$, and $\odot$ corresponds to an element-wise (Hadamard) product between matrices.

\subsection{Proposed FC Filtering-based Iterative PAPR Reduction}
\label{subsec:FC_with_it_PAPR}

\subsubsection{Baseline FC-F-OFDM Processing \cite{fc2}}

\begin{figure}[tb]
	\centering
	\includegraphics[width=0.95\columnwidth]{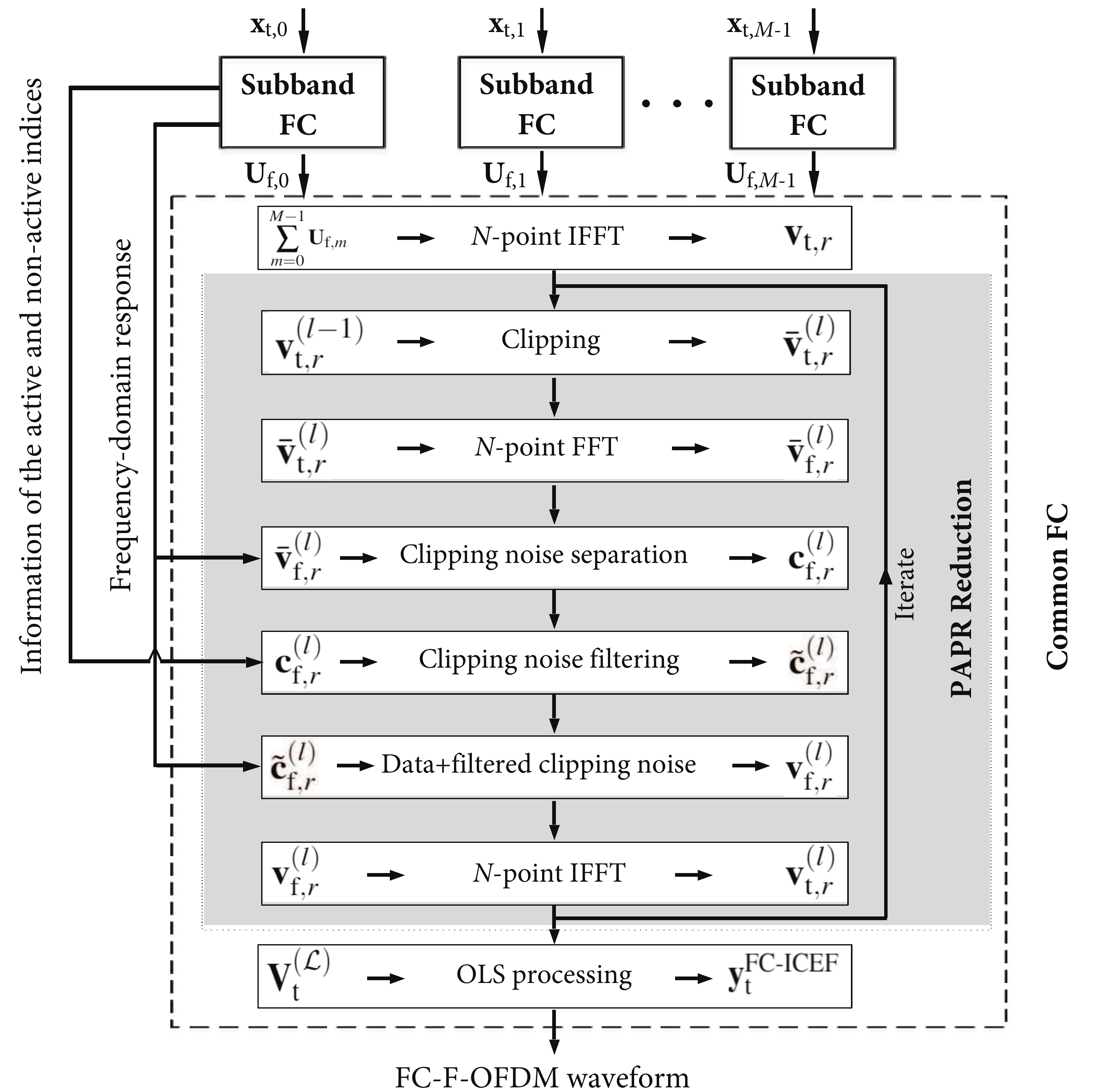}
	\caption{Block diagram of the proposed FC processing based PAPR reduction applied within the ``PAPR and INI Reduction'' phase as illustrated in Fig. \ref{fig:PAPRandINI}.}
	\vspace{-2mm}
	\label{fig:FCICFblock}
\end{figure}

The first step of the FC filtering is the subband-specific processing done within the ``Subband FC" phase shown in Fig. \ref{fig:FCICFblock}. In this phase, each of the $M$ subband signals is segmented into $R_\textrm{FC}$ overlapping FC processing blocks of length $L_m$ with overlap of $L_{\text{O},m}$ samples, and these blocks are collected into an $L_m\times R_\textrm{FC}$ matrix $\mathbf{B}_{\textrm{t},m}$. The length of the non-overlapping part is $L_{\text{S},m}=L_m-L_{\text{O},m}$ and the overlap factor is $\lambda=1-L_{\text{S},m}/L_m$. Following \cite{fc2}, the subband-specific processing done within ``Subband FC" phase can be expressed as
\begin{align}
 \textbf{U}_{\textrm{f},m} = \textbf{M}_{m} \textbf{D}_{m} \textbf{P}_{m}^{(L_m/2)} \textbf{W}_{\textrm{FC},m} \mathbf{B}_{\textrm{t},m} \Theta_m, 
\label{umat}
\end{align}
where $\textbf{W}_{\textrm{FC},m}$ represents the $L_m\times L_m$ DFT matrix, $\textbf{P}_{m}^{(L_m/2)}$ represents the DFT-shift matrix obtained by cyclically shifting left the $L_m\times L_m$ identity matrix by $L_m/2$ positions, and $\textbf{D}_{m}$ denotes the $L_m\times L_m$ diagonal matrix that contains the frequency-domain window weights $\mathbf{d}_m$ of the $m\textrm{th}$ subband on the main diagonal. Here, the frequency-domain window $\mathbf{d}_m$ corresponds to the DFT of a finite-length linear filter impulse response. Matrix $\textbf{M}_{m}$ of size $N\times L_m$ maps input's $L_m$ frequency-domain bins to output signal's frequency-domain bins $(c_m-\lceil L_m/2\rceil+b)_N$ for $b=0,1,\ldots,L_m-1$. Here, $c_m$ and $(\cdot)_N$ represent the center of the $m\textrm{th}$ subband and modulo-$N$ operation, respectively. Furthermore, to provide phase continuity between consecutive processing blocks, the diagonal matrix $\Theta_m$ of size $R_\textrm{FC} \times R_\textrm{FC}$ rotates the phase of the $r$th block by $[\Theta_m]_{r,r}=\exp(j2\pi r\theta_m)$, where $\theta_m={c_m L_{\text{S},m}}/{L_m}$.

In the ``Common FC" phase, the frequency-domain windowed subband signals $\mathbf{U}_{\textrm{f},m}$ are combined to obtain the high-rate time-domain FC processing blocks denoted as
\begin{align}
    \textbf{V}_\textrm{t}  = [\textbf{v}_{\textrm{t},0}, \textbf{v}_{\textrm{t},1}, .\dots, \textbf{v}_{\textrm{t},{R_\textrm{FC}-1}}] = \textbf{W}_{\textrm{FC}}^{-1}\sum_{m=0}^{M-1}\textbf{U}_{\textrm{f},m},
    \label{requ}
\end{align}
where $\textbf{W}_{\textrm{FC}}^{-1}$ represents the $N\times N$ IDFT matrix and
the FC processing block-specific time-domain output vector is denoted as $\textbf{v}_{\textrm{t},r}$, where $r \in \{ 0,1, \dots, R_\textrm{FC}-1 \}$. The filtered mixed-numerology FC-F-OFDM waveform is obtained as $\mathbf{y}_{\textrm{t}}^\textrm{FC} = \VEC(\Upsilon_{N}\textbf{V}_\textrm{t})$, where the $N_\text{S}\times N$ selection matrix $\Upsilon_N$ selects the required $N_\text{S}=N L_{\text{S},m}/L_m$ non-overlapping samples corresponding to overlap-and-save (OLS) processing. These steps are depicted as the first and last subblocks of the ``Common FC" phase as illustrated in Fig. \ref{fig:FCICFblock}, whereas the other subblocks relate to the proposed PAPR reduction scheme described in more detail in the following.

\subsubsection{Proposed FC-filtered ICEF (FC-ICEF)}

In the proposed method, the ICEF-like mechanism is embedded into ``Common FC" phase and by exploiting the information about active bin indices in the filtering phase of FC processing, efficient PAPR reduction is achieved without significant degradation in FC-F-OFDM's out-of-band (OOB) emission performance. The proposed technique is based on iteratively clipping the time-domain wideband FC processing blocks $\mathbf{v}^{(l-1)}_{\textrm{t},r}$, instead of OFDM symbols, for which we have the knowledge of the active and non-active frequency parts. Thus, in the following, ``subcarriers" refer terminology-wise to the structure of the OFDM symbols and ``frequency bins" refer to the frequency content of the FC processing blocks.

Specifically, as illustrated in Fig. \ref{fig:FCICFblock}, iterations start with the clipping operation, applied to FC processing blocks, such that $\bar{\mathbf{v}}^{(l)}_{\textrm{t},r}$ represents the clipped version of $ \mathbf{v}^{(l-1)}_{\textrm{t},r}$. Then, the clipped signal is converted into frequency domain by a DFT of size $N$, that results in $\bar{\mathbf{v}}_{\textrm{f},r}^{(l)}$. Similar to the ordinary ICEF \cite{icef}, the clipping noise is separated from the clipped data-bearing signal and processed separately. Accordingly, the clipping noise at the $l\textrm{th}$ iteration can be obtained in frequency domain as 
\begin{align}
\mathbf{c}^{(l)}_{\textrm{f},r} = \bar{\mathbf{v}}^{(l)}_{\textrm{f},r} - \mathbf{v}_{\textrm{f},r},
\label{equc}
\end{align} 
where $\mathbf{v}_{\textrm{f},r}$ is the frequency-domain representation of the $r\textrm{th}$ FC processing block without clipping noise defined in (\ref{requ}).

\begin{figure*} [tb]
	\centering
	\subfloat{\label{PAPRdists}\includegraphics[width=0.3\linewidth]{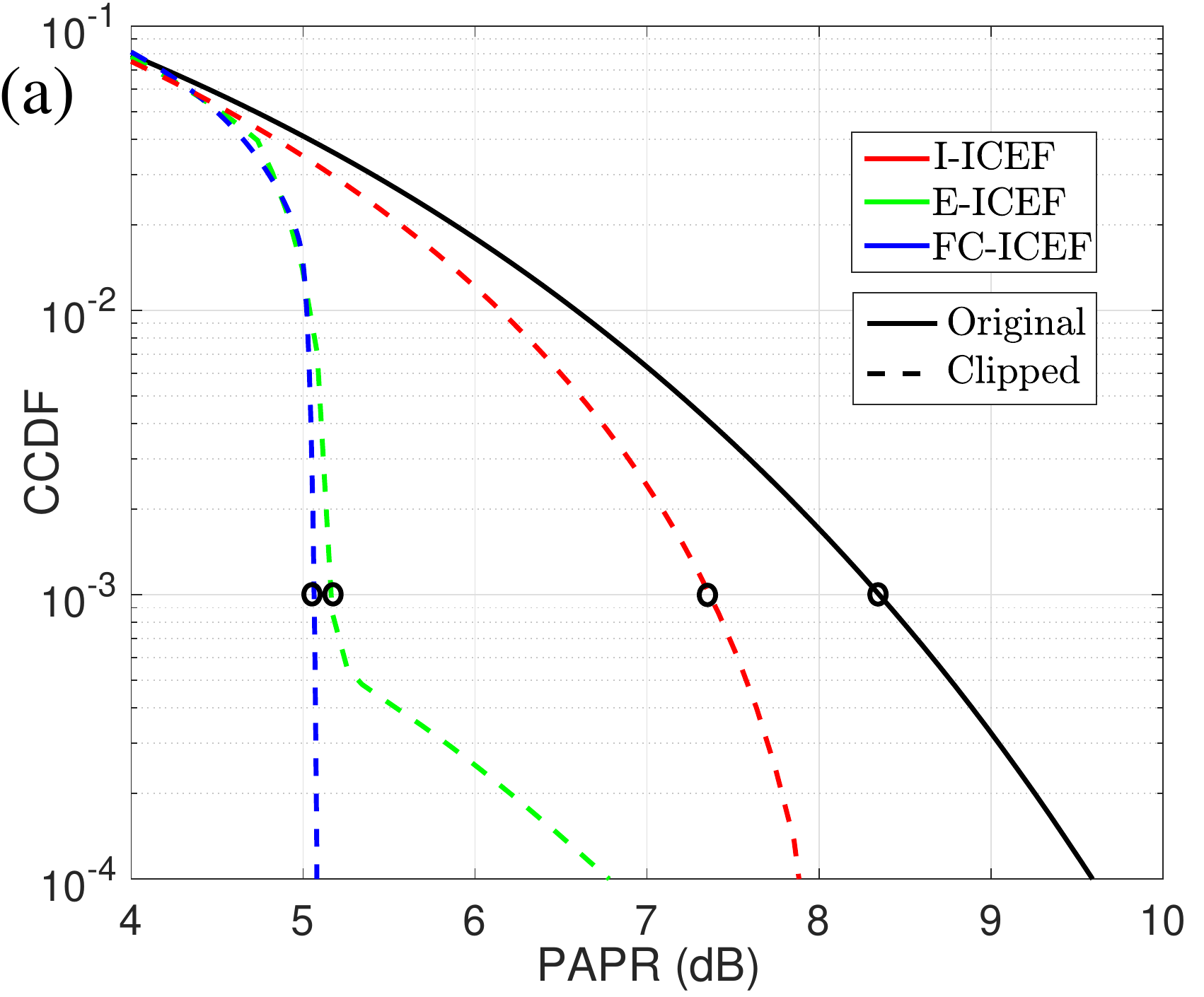}}
	\subfloat{\label{FCMSE}\includegraphics[width=0.35\linewidth]{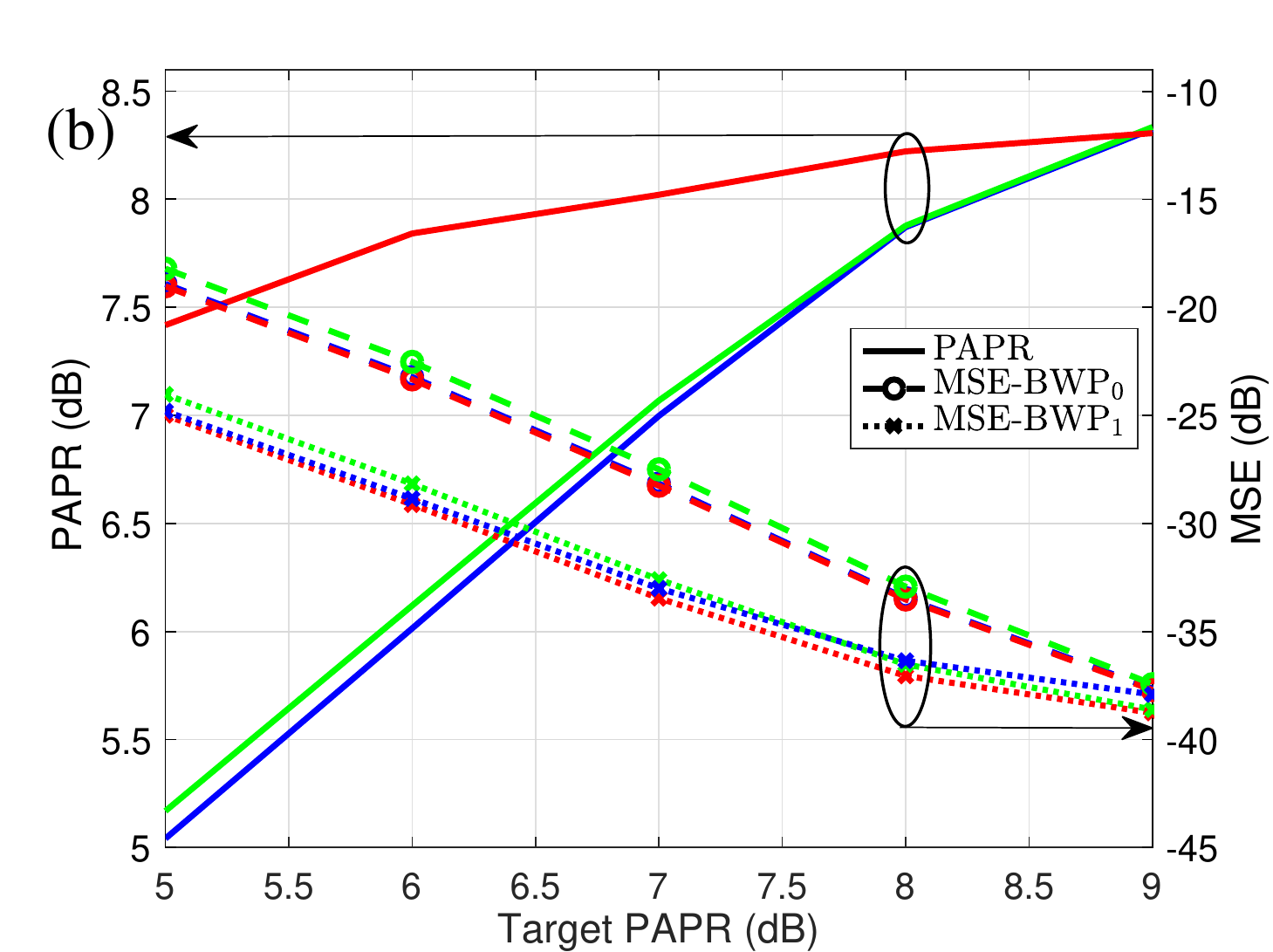}}
	\subfloat{\label{OOB}\includegraphics[width=0.35\linewidth]{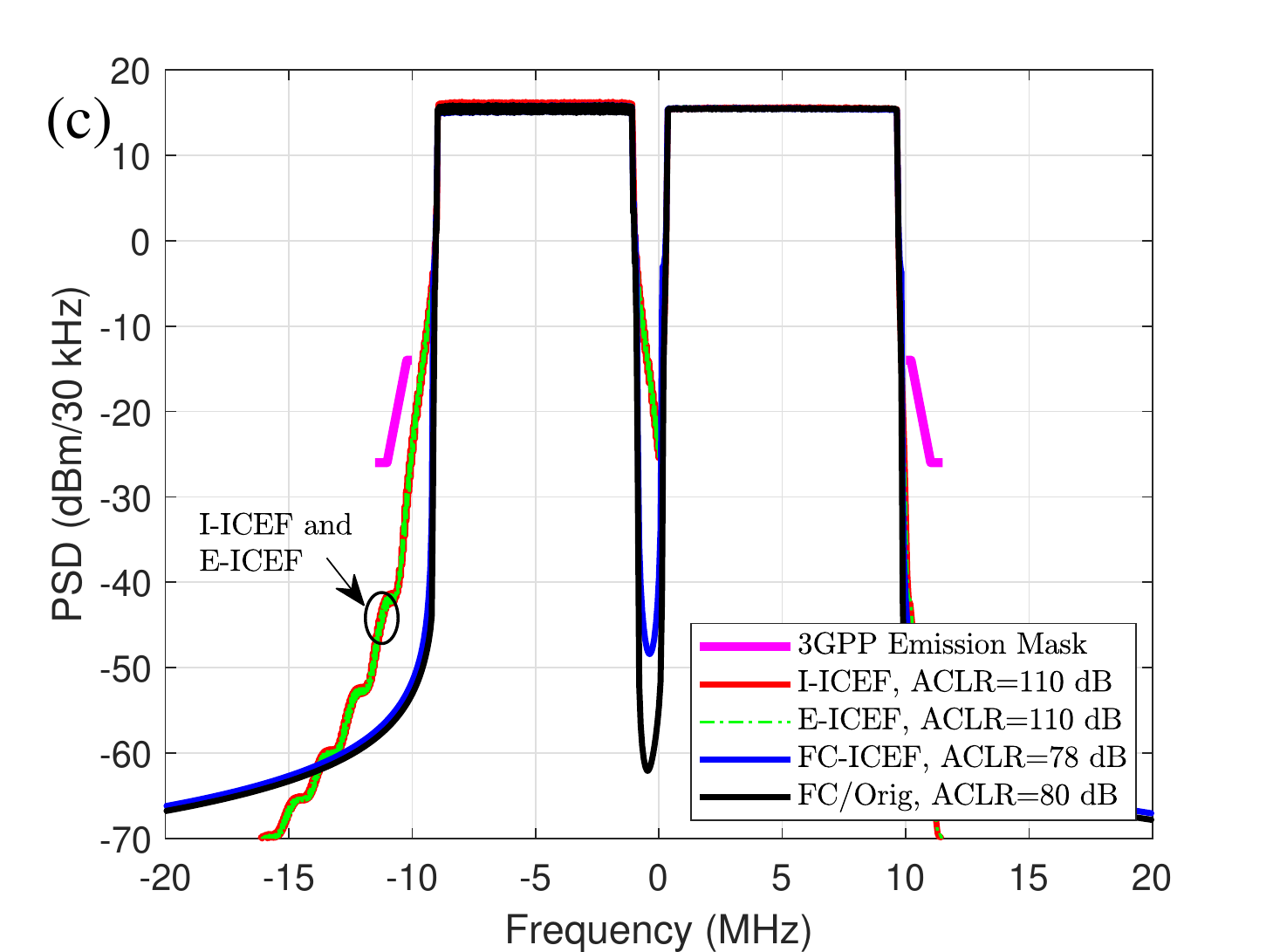}}
	\caption{Evaluation results and comparison of FC-ICEF, E-ICEF, and I-ICEF. In (a), PAPR distributions are shown for PAPR target of \SI{5}{dB}. CCDF level of 0.1$\%$ is highlighted with black circles. For different PAPR target levels, the achieved PAPR at 0.1$\%$ probability level and the corresponding MSE results are shown in (b). In (c), for PAPR clipping target level of \SI{5}{dB}, PSDs of each case are shown with the corresponding emission mask for \SI{30}{kHz} measurement bandwidth \cite{3GPPTS38104} while ACLR results for all cases are also given in the legend.
	\label{fig:papr_mse_oob} 
	}
\end{figure*}

Next, frequency-selective filtering is applied to suppress the clipping noise outside the FC frequency bins where clipping noise is allowed. The used clipping error filter is defined as
\begin{align}
h^{\textrm{FC}}_{\textrm{f},r}[n] = \begin{cases}
1, & \text{if} \; n\in\mathcal{K}_{E},\\
0, & \text{if} \; n\in\mathcal{K_F} \cup \mathcal{K}_{null},
\end{cases}
\label{filter}
\end{align}
where $n \in \{0, 1, \ldots, N-1\}$, and $\mathcal{K}_E$ and $\mathcal{K_F}$ represent the mutually exclusive sets of FC frequency bins within the channel bandwidth where clipping noise is allowed and not allowed, respectively, while $\mathcal{K}_{null}$ is the set of FC frequency bins outside the used channel bandwidth, and $\mathcal{K}_{active}=\mathcal{K}_E \cup \mathcal{K_F}$. 

These index sets are created by exploiting the subband-specific FC filter information about passband, transition band, and stopband FC frequency bin indices. In the following examples, $\mathcal{K}_{active}$ covers the NR \SI{20}{MHz} channel bandwidth, and $\mathcal{K}_E$ is configured as the set that contains passband and transition band FC frequency bin indices for all $M$ subbands. Remaining indices of the $\mathcal{K}_{active}$ are contained in $\mathcal{K_F}$. Consequently, the proposed ICEF filter efficiently masks clipping noise over utilized {subbands}. It should also be noted, that using non-trivial weights in the clipping noise filter $\mathbf{h}_{\textrm{f},r}^{\textrm{FC}}$ allows to achieve QoS specific error levels per subband, but the detailed analysis of such approach is outside the scope of this letter.

After (\ref{equc}), the obtained frequency-domain clipping noise is filtered and added back to the original non-clipped signal, which can be expressed as 
\begin{align}
\mathbf{v}^{(l)}_{\textrm{f},r} 
= \mathbf{v}_{\textrm{f},r}+\mathbf{h}^{\textrm{FC}}_{\textrm{f},r} \odot \mathbf{c}^{(l)}_{\textrm{f},r} 
= \mathbf{v}_{\textrm{f},r}+\tilde{\mathbf{c}}^{(l)}_{\textrm{f},r}.   
\label{last}
\end{align}
The PAPR-reduced FC processing block is then converted to time domain through IDFT of size $N$, and is denoted as $\mathbf{v}_{\textrm{t},r}^{(l)}$. Finally, after $\mathcal{L}$ iterations or reaching the target PAPR level, the FC-F-OFDM signal with reduced PAPR is obtained after OLS processing, denoted as  $\mathbf{y}_{\textrm{t}}^\textrm{FC-ICEF}=\VEC(\Upsilon_{N}{\mathbf{V}}^{(\mathcal{L})}_{\textrm{t}})$. 

It is shortly noted that in terms of complexity, the I-ICEF reference method has the lowest PAPR reduction complexity. In the case of E-ICEF, the complexity is increased mainly due to the calculation of the subband specific INI components $\mathbf{z}^{(l-1)}_{\textrm{f},m,s}$ and creating the aggregated signal in each iteration. In the case of FC-ICEF, the computational complexity per FC processing block per iteration corresponds to the complexity of the original ICEF algorithm \cite{icef}. In FC filtering, the number of FC processing blocks varies depending on the used overlap factor and bin spacing. With the evaluated parameters, shown in Table~\ref{table:parameters}, and when compared to I-ICEF, the number of real multiplications is increased by 45\% or 18\% with E-ICEF or FC-ICEF, and the number of real additions is increased by 50\% or 16\% with E-ICEF or FC-ICEF, respectively. In these comparisons, only the PAPR reduction complexity is considered, whereas the complexity of WOLA and FC processing is analyzed and compared in \cite{fc2,2018:J:Levanen:Transparent5G}.

\section{Numerical Results and Analysis}

In this section, the proposed solutions are analyzed and compared through numerical evaluations. Their performance is measured in terms of MSE, PAPR, OOB emission, and adjacent-channel-leakage-ratio (ACLR). In particular, MSE is measured by following the 3GPP measurement procedure and error-vector magnitude (EVM) window lengths defined in \cite{3GPPTS38104}. Similarly, OOB emissions and ACLR are evaluated by following the same specification, and the corresponding emission masks are also included in the results. The assumed 5G NR channel bandwidth is \SI{20}{MHz}, and it is divided into two BWPs, each configured as a \SI{10}{MHz} subband with different SCSs, by following the 5G NR radio interface numerology defined in \cite{3GPPTS38104}. Moreover, the BWPs are centered at \SI{-5}{MHz} and \SI{5}{MHz} positions, relative to the overall 20 MHz channel's center frequency. The main physical-layer and evaluation parameters are shown in Table~\ref{table:parameters}.

\begin{table}[t]
  \setlength{\tabcolsep}{2.8pt}
    \renewcommand{\arraystretch}{1.3}
    \footnotesize
    \centering
    \caption{The main evaluation parameters}
    \begin{tabular}{|l|c|}
    \hline
    \multicolumn{1}{|c|}{\textbf{Parameter}} & \multicolumn{1}{c|}{\textbf{Value}}\\
		\hline 
        Physical resource block (PRB) size & 12 subcarriers \\ \hline
		Subcarrier spacing (SCS) for $\textrm{BWP}_m$, $m \in \{0,1\}$ & \{15, 60\}~kHz \\ \hline
		Number of PRBs & \{52, 11\} \\ \hline
		Nominal OFDM transform size ($L_{\textrm{OFDM},m}$) & \{2048, 512\} \\ \hline
		Number of transmitted OFDM symbols ($S_m$) & \{8192, 32768\} \\ \hline
		Oversampling factor ($N_\textrm{ov}$) & 4  \\ \hline
		Modulation order & \{QPSK, 64-QAM\} \\ \hline
		Maximum number of (I/E/FC)-ICEF iterations & 20 \\ \hline 
		FC-F-OFDM: Nominal inverse transform size ($N_\textrm{nom}$) & 2048 \\ 
		FC-F-OFDM: Bin spacing & \SI{15}{kHz} \\
        FC-F-OFDM: Overlap factor  & 1/2 \\
        FC-F-OFDM: Number of transition band bins & 12 \\ 
        FC-F-OFDM: Transition band response & Raised-cosine \\ \hline
        WOLA: Nominal window length ($L_{\textrm{WOLA},m}$) & $1.7L_{\textrm{CP},m}+L_{\textrm{OFDM},m}$ \\ \hline
    \end{tabular}
    \label{table:parameters}
\end{table}

The sample-wise PAPR of the $n\textrm{th}$ sample of the output signal $\mathbf{y}^{(l)}_{\textrm{t}}$ is defined and quantified as
\begin{align}
    \text{PAPR}\big(y^{(l)}_{\textrm{t}}[n]\big) = \frac{|y^{(l)}_{\textrm{t}}[n]|^2}{\frac{1}{N_{\textrm{y}}} \sum_{n=0}^{N_{\textrm{y}}-1}|y^{(l)}_{\textrm{t}}[n]|^2 }, 
    \label{PAPRcalcu}
\end{align}
where $N_{\textrm{y}}$ denotes the length of  $\mathbf{y}^{(l)}_{\textrm{t}}$. To achieve accurate representation of the PAPR complementary cumulative distribution function (CCDF), an oversampling factor of $N_\textrm{ov}=4$ is used, similar to \cite{J:2002:Ochiai:PerformanceClippedOFDM}. With FC-ICEF, the FC processing allows to effectively increase the sampling rate by a factor of $I_m=N/L_m=N_\text{S}/L_{\text{S},m}$, where sampling rate can be changed flexibly by configuring forward and inverse transform lengths accordingly. In the used configuration, inverse transform size of $N=N_\textrm{ov}N_\textrm{nom}=8192$ is used to achieve the desired oversampling rate. Furthermore, in the same case, the CP-OFDM signals are generated by using the nominal OFDM transform sizes \cite{3GPPTS38104}. With I-ICEF and E-ICEF, in turn, the oversampling factor is directly included in the OFDM symbol's IDFT, and the used transform sizes correspond to $L_{\textrm{OFDM},0}=8192$ and $L_{\textrm{OFDM},1}=2048$ for $\textrm{BWP}_0$ and $\textrm{BWP}_1$ carrying OFDM signals with SCSs of \SI{15}{kHz} and \SI{60}{kHz}, respectively.
 
The achievable PAPR performance is evaluated for different PAPR targets, ranging from \SI{5}{dB} to \SI{9}{dB}. As transition band weights, 12 frequency bins following the well-known RC response are considered in FC processing, which is suitable for the considered scenario and provides a good trade-off between MSE and OOB emission performance. With WOLA, in order to fulfill the OOB emission requirements \cite{3GPPTS38104}, an appropriate window length was adopted,
corresponding to an overlap of about 35\% of the CP length between adjacent RC-windowed cyclically extended OFDM symbols.

First, the PAPR performance of different methods is compared and corresponding results are shown in Fig.~\ref{fig:papr_mse_oob}(a). In this case, the PAPR target of \SI{5}{dB} is used and the PAPR distribution of the original mixed-numerology signal is also given as a reference. It is clear that the proposed FC-ICEF provides very good performance, since the target PAPR level is almost achieved with \SI{0.04}{dB} offset at CCDF probability level of $0.1\%$. Furthermore, the E-ICEF and I-ICEF methods provide approximately \SI{0.13}{dB} and \SI{2.38}{dB} worse performance results, respectively. In the case of I-ICEF, the final combination of the separately PAPR limited signals results in poor PAPR reduction on the aggregated signal. In the E-ICEF case, in turn, the mixed-numerology interference during the ICEF iterations is effectively attenuated leading to very good PAPR reduction performance, close to that of FC-ICEF. It can also be observed that with E-ICEF, there is still some residual crosstalk between the different numerologies leading to increasing PAPR values below the reference CCDF probability level of $0.1\%$. 

Next, the PAPR performance at $0.1\%$ probability level as well as the average passband MSE for both BWPs are provided in Fig. \ref{fig:papr_mse_oob} (b), for different PAPR target levels. Again, very poor PAPR reduction performance of the I-ICEF is clearly visible. For the FC-ICEF, the targeted PAPR levels are accurately reached, and the E-ICEF also nearly achieves these, losing only some \SI{0.2}{dB} at PAPR target levels of \SI{5}{dB} and \SI{6}{dB}. The MSE of all methods saturates approximately to the level of \SI{-39}{dB} due to the reference CP-OFDM receiver, which is known to introduce some INI \cite{fc2,2018:J:Levanen:Transparent5G}. All methods provide similar MSE performance over the evaluated PAPR target value range and, interestingly, the BWP$_1$ with larger subcarrier spacing is least affected by the clipping noise. The E-ICEF with WOLA is not able to compensate the INI exactly as efficiently as the FC-ICEF, therefore, it has slightly higher MSE over the evaluated target PAPR range. For practical reference, the MSE requirements of QPSK or 64-QAM modulations are \SI{-15}{dB} or \SI{-22}{dB}, respectively \cite{3GPPTS38104}. These MSE thresholds can thus be clearly satisfied with all methods, even with PAPR target level as low as \SI{5}{dB}, noting that BWP$_0$ and BWP$_1$ carry QPSK and 64-QAM data, respectively. 

Finally, the performance of the evaluated methods is measured in terms of OOB emission and ACLR. The output signal power spectral densities (PSDs) of the different schemes are shown in Fig. \ref{fig:papr_mse_oob} (c). In general, since the ICEF-like methods null the clipping noise on the OOB region, PAPR reduction does not cause any larger additional OOB emissions. Given the clipping noise filter's definition in Section \ref{subsec:FC_with_it_PAPR}, FC-ICEF is able to also effectively suppress the clipping noise between the BWPs and at guard bands within the channel bandwidth, allowing for improved coexistence. With WOLA-based methods, the attenuation is better further away from the active bands, leading to higher ACLR values, stemming from the relatively large window length. The FC-ICEF also provides a very good ACLR performance, which is \SI{78}{dB} in this case. In order to better judge and quantify the quality of the FC-ICEF based signal, the PSD and ACLR results of the original FC-F-OFDM waveform, with no PAPR reduction, are also shown for reference. As can be observed, the presented PAPR reduction approach reduces the ACLR of the FC filtered signal only by some \SI{2}{dB}.

\section{Conclusions}
\label{sec:conclusions}

Two novel schemes for reducing the PAPR of a mixed-numerology CP-OFDM signal were presented. The first scheme is an enhanced ICEF algorithm operating on OFDM symbol level that cancels the inter-numerology interference along the PAPR reduction, making it compatible with traditional WOLA-based spectral shaping. The second scheme, embedding ICEF-like processing with fast-convolution (FC) filtering, operates on the FC processing blocks instead of OFDM symbols. This approach thus allows limiting the PAPR of any kind of input signal, as it relies on the block-based processing inherent to FC processing, which is independent of the input signal. The performance of both schemes was evaluated in terms of PAPR reduction, passband MSE, and OOB emission levels, showing excellent performance in mixed-numerology operation. Both presented schemes allow to efficiently reduce the PAPR of mixed-numerology signals, which is crucial for the new 5G NR radio interface and for communications beyond 5G.

\bibliographystyle{IEEEtran}
\bibliography{IEEEabrv,IEEEref}

\end{document}